\documentclass[fleqn,10pt]{wlscirep}
\usepackage[utf8]{inputenc}
\usepackage[T1]{fontenc}
\usepackage{bm}
\title{Another view on Gilbert damping in two-dimensional ferromagnets}

\usepackage{color}

\author[1]{Anastasiia A. Pervishko}
\author[1,2]{Mikhail I. Baglai}
\author[2,3]{Olle Eriksson}
\author[1]{Dmitry Yudin}
\affil[1]{ITMO University, Saint Petersburg 197101, Russia}
\affil[2]{Department of Physics and Astronomy, Uppsala University, Box 516, SE-75 121 Uppsala, Sweden}
\affil[3]{School of Science and Technology, \"{O}rebro University, SE-701 82 \"{O}rebro, Sweden}

\begin{abstract}
A keen interest towards technological implications of spin-orbit driven magnetization dynamics requests a proper theoretical description, especially in the context of a microscopic framework, to be developed. Indeed, magnetization dynamics is so far approached within Landau-Lifshitz-Gilbert equation which characterizes torques on magnetization on purely phenomenological grounds. Particularly, spin-orbit coupling does not respect spin conservation, leading thus to angular momentum transfer to lattice and damping as a result. This mechanism is accounted by the Gilbert damping torque which describes relaxation of the magnetization to equilibrium. In this study we work out a microscopic Kubo-St\v{r}eda formula for the components of the Gilbert damping tensor and apply the elaborated formalism to a two-dimensional Rashba ferromagnet in the weak disorder limit. We show that an exact analytical expression corresponding to the Gilbert damping parameter manifests linear dependence on the scattering rate and retains the constant value up to room temperature when no vibrational degrees of freedom are present in the system. We argue that the methodology developed in this paper can be safely applied to bilayers made of non- and ferromagnetic metals, e.g., CoPt. 
\end{abstract}
\begin{document}

\flushbottom
\maketitle

\thispagestyle{empty}

\section*{Introduction}

In spite of being a mature field of research, studying magnetism and spin-dependent phenomena in solids still remains one of the most exciting area in modern condensed matter physics. In fact, enormous progress in technological development over the last few decades is mainly held by the achievements in spintronics and related fields \cite{Zutic2004,Bader2010,MacDonald2011,Koopmans2011,Gomonay2014,Jungwirth2016,Duine2018,Zelezny2018,Nemec2018,Smejkal2018,Baltz2018}. However the theoretical description of magnetization dynamics is at best accomplished on the level of Landau-Lifshitz-Gilbert (LLG) equation that characterizes torques on the magnetization. In essence, this equation describes the precession of the magnetization, $\bm{m}(\bm{r},t)$, about the effective magnetic field, $\bm{H}_\mathrm{eff}(\bm{r},t)$, created by the localized moments in magnetic materials, and its relaxation to equilibrium. The latter, known as the Gilbert damping torque \cite{Gilbert2004}, was originally captured in the form $\alpha\bm{m}\times\partial_t\bm{m}$, where the parameter $\alpha$ determines the relaxation strength, and it was recently shown to originate from a systematic non-relativistic expansion of the Dirac equation \cite{Hickey2009}. Thus, a proper microscopic determination of the damping parameter $\alpha$ (or, the damping tensor in a broad sense) is pivotal to correctly simulate dynamics of magnetic structures for the use in magnetic storage devices \cite{Sharma2017}. 

From an experimental viewpoint, the Gilbert damping parameter can be extracted from ferromagnetic resonance linewidth measurements \cite{Scheck2007,Woltersdorf2009,Zhao2016} or established via time-resolved magneto-optical Kerr effect \cite{Iihama2014,Capua2015}. In addition, it was clearly demonstrated that in bilayer systems made of a nonmagnetic metal (NM) and a ferromagnet material (FM) the Gilbert damping is drastically enhanced as compared to bulk FMs \cite{Heinrich1987,Platow1998,Urban2001,Mizukami2002,He2013}. A strong magnetocrystalline anisotropy, present in  CoNi, CoPd, or CoPt, hints unambiguously for spin-orbit origin of the intrinsic damping. A first theoretical attempt to explain the Gilbert damping enhancement was made in terms of $sd$ exchange model in Ref.~\cite{Berger1996}. Within this simple model, magnetic moments associated with FM layer transfer angular momentum via interface and finally dissipate. Linear response theory has been further developed within free electrons model \cite{Simanek2003,Mills2003}, while the approach based on scattering matrix analysis has been presented in Refs.~\cite{Tserkovnyak2002a,Tserkovnyak2002b}. In the latter scenario spin pumping from FM to NM results in either backscattering of magnetic moments to the FM layer or their further relaxation in the NM. Furthermore, the alternative method to the evaluation of the damping torque, especially in regard of first-principles calculations, employs torque-correlation technique within the breathing Fermi surface model \cite{Kambersky2007}. While a direct estimation of spin-relaxation torque from microscopic theory \cite{Nakabayashi2010}, or from spin-wave spectrum, obtained on the basis of transverse magnetic field susceptibility \cite{Costa2010,Santos2013}, are also possible. It is worth mentioning that the results of first-principles calculations within torque-correlation model \cite{Gilmore2007,Garate2009,Thonig2014,Schoen2016,Thonig2018} and linear response formalism \cite{Ebert2011,Mankovsky2013} reveal good agreement with experimental data for itinerant FMs such as Fe, Co, or Ni and binary alloys.

Last but not least, an intensified interest towards microscopic foundations of the Gilbert parameter $\alpha$ is mainly attributed to the role the damping torque is known to play in magnetization reversal \cite{Ralph2008}. In particular, according to the breathing Fermi surface model the damping stems from variations of single-particle energies and consequently a change of the Fermi surface shape depending on spin orientation. Without granting any deep insight into the microscopic picture, this model suggests that the damping rate depends linearly on the electron-hole pairs lifetime which are created near the Fermi surface by magnetization precession. In this paper we propose an alternative derivation of the Gilbert damping tensor within a mean-field approach according to which we consider itinerant subsystem in the presence of nonequilibrium classical field $\bm{m}(\bm{r},t)$. Subject to the function $\bm{m}(\bm{r},t)$ is sufficiently smooth and slow on the scales determined by conduction electrons mean free path and scattering rate, the induced nonlocal spin polarization can be approached within a linear response, thus providing the damping parameter due to the itinerant subsystem. In the following, we provide the derivation of a Kubo-St\v{r}eda formula for the components of the Gilbert damping tensor and illustrate our approach for a two-dimensional Rashba ferromagnet, that can be modeled by the interface between NM and FM layers. We argue that our theory can be further applied to identify properly the tensorial structure of the Gilbert damping for more complicated model systems and real materials.

\section*{Microscopic framework}

\begin{figure}
\includegraphics[width=0.4\textwidth]{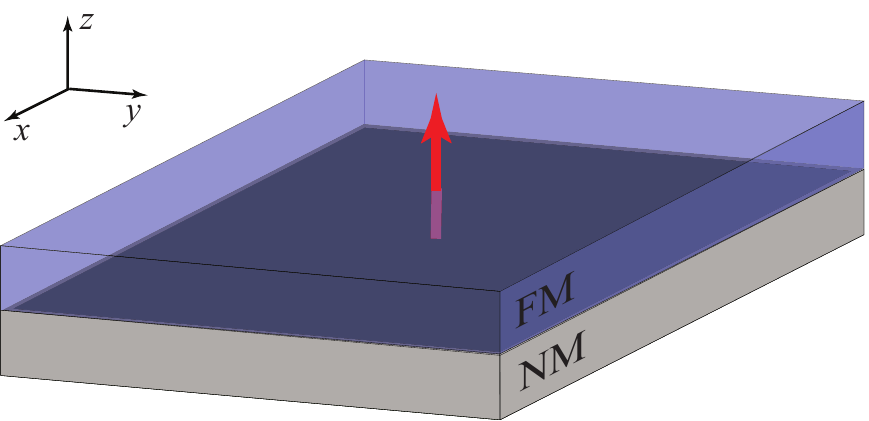}
\caption{Schematic representation of the model system: the electrons at the interface of a bilayer, composed of a ferromagnetic (FM) and a nonmagnetic metal (NM) material, are well described by the Hamiltonian (\ref{hami}). We assume the magnetization of FM layer depicted by the red arrow is aligned along the $z$ axis.}\label{fig:fig1}
\end{figure}

Consider a heterostructure made of NM with strong spin-orbit interaction covered by FM layer as shown in Fig.~\ref{fig:fig1}, e.g., CoPt. In general FMs belong to the class of strongly correlated systems with partially filled $d$ or $f$ orbitals which are responsible for the formation of localized magnetic moments. The latter can be described in terms of a vector field $\bm{m}(\bm{r},t)$ referred to as magnetization, that in comparison to electronic time and length scales slowly varies and interacts with an itinerant subsystem. At the interface (see Fig.~\ref{fig:fig1}) the conduction electrons of NM interact with the localized magnetic moments of FM via a certain type of exchange coupling, $sd$ exchange interaction, so that the Hamiltonian can be written as
\begin{equation}\label{hami}
h=\frac{p^2}{2m}+\alpha\left(\bm{\sigma}\times\bm{p}\right)_z+\bm{\sigma}\cdot\bm{M}(\bm{r},t)+U(\bm{r}),
\end{equation}
where first two terms correspond to the Hamiltonian of conduction electrons, on condition that the two-dimensional momentum $\bm{p}=(p_x,p_y)=p(\cos\varphi,\sin\varphi)$ specifies electronic states, $m$ is the free electron mass, $\alpha$ stands for spin-orbit coupling strength, while $\bm{\sigma}=(\sigma_x,\sigma_y,\sigma_z)$ is the vector of Pauli matrices. The third term in (\ref{hami}) is responsible for $sd$ exchange interaction with the exchange field $\bm{M}(\bm{r},t)=\Delta\,\bm{m}(\bm{r},t)$ aligned in the direction of magnetization and $\Delta$ denoting $sd$ exchange coupling strength. We have also included the Gaussian disorder, the last term in Eq.~(\ref{hami}), which represents a series of point-like defects, or scatterers, $\langle U(\bm{r})U(\bm{r}')\rangle=(m\tau)^{-1}\delta(\bm{r}-\bm{r}')$ with the scattering rate $\tau$ (we set $\hbar=1$ throughout the calculations and recover it for the final results).

Subject to the norm of the vector $|\bm{m}(\bm{r},t)|=1$ remains fixed, the magnetization, in broad terms, evolves according to (see, e.g., Ref.~\cite{Ado2017}),
\begin{equation}\label{dyneq}
\partial_t\bm{m}=\bm{f}\times\bm{m}=\gamma\bm{H}_\mathrm{eff}\times\bm{m}+\chi\bm{s}\times\bm{m},
\end{equation}
where $\bm{f}$ corresponds to so-called spin torques. The first term in $\bm{f}$ describes precession around the effective magnetic field $\bm{H}_\mathrm{eff}$ created by the localized moments of FM, whereas the second term in (\ref{dyneq}) is determined by nonequilibrium spin density of conduction electrons of NM at the interface, $\bm{s}(\bm{r},t)$. It is worth mentioning that in Eq.~(\ref{dyneq}) the parameter $\gamma$ is the gyromagnetic ratio, while $\chi=(g\mu_B/\hbar)^2\mu_0/d$ is related to the electron $g-$factor ($g=2$), the thickness of a nonmagnetic layer $d$, with $\mu_B$ and $\mu_0$ standing for Bohr magneton and vacuum permeability respectively.
Knowing the lesser Green's function, $G^<(\bm{r}t;\bm{r}t)$, one can easily evaluate nonequilibrium spin density of conduction electrons induced by slow variation of magnetization orientation,
\begin{equation}\label{spin}
s_\mu(\bm{r},t)=-\frac{i}{2}\mathrm{Tr}\left[\sigma_\mu G^<(\bm{r}t;\bm{r}t)\right]=Q_{\mu\nu}\partial_t m_\nu+\ldots,
\end{equation}
where summation over repeated indexes is assumed ($\mu,\nu=x,y,z$). The lesser Green's function of conduction electrons can be represented as $G^<=\left(G^K-G^R+G^A\right)/2$, where $G^K$, $G^R$, $G^A$ are Keldysh, retarded, and advanced Green's functions respectively.   

\subsection*{Kubo-St\v{r}eda formula} We further proceed with evaluating $Q_{\mu\nu}$ in Eq.~(\ref{spin}) that describes the contribution to the Gilbert damping due to conduction electrons. In the Hamiltonian (\ref{hami}) we assume slow dynamics of the magnetization, such that approximation $\bm{M}(\bm{r},t)\approx\bm{M}+(t-t_0)\,\partial_t\bm{M}$ with $\bm{M}=\bm{M}(\bm{r},t_0)$ is supposed to be hold with high accuracy,
\begin{equation}\label{inham}
H=\frac{p^2}{2m}+\alpha\left(\bm{\sigma}\times\bm{p}\right)_z+\bm{\sigma}\cdot\bm{M}+U(\bm{r})+(t-t_0)\,\bm{\sigma}\cdot\partial_t\bm{M},
\end{equation}
where first four terms in the right hand side of Eq.~(\ref{inham}) can be grouped into the Hamiltonian of a bare system, $H_0$, which coincides with that of Eq.~(\ref{hami}), provided by the static magnetization configuration $\bm{M}$. In addition, the expression (\ref{inham}) includes the time-dependent term $V(t)$ explicitly, as the last term. In the following analysis we deal with this in a perturbative manner. In particular, the first order correction to the Green's function of a bare system induced by $V(t)$ is,
\begin{equation}
\delta G(t_1,t_2)=\int_{C_K}dt\int\frac{d^2p}{(2\pi)^2}\,g_{\bm{p}}(t_1,t)V(t)g_{\bm{p}}(t,t_2),
\end{equation}
where the integral in time domain is taken along a Keldysh contour, while $g_{\bm{p}}(t_1,t_2)=g_{\bm{p}}(t_1-t_2)$ [the latter accounts for the fact that in equilibrium correlation functions are determined by the relative time $t_1-t_2$] stands for the Green's function of the bare system with the Hamiltonian $H_0$ in momentum representation. In particular, for the lesser Green's function at coinciding time arguments $t_1=t_2\equiv t_0$, which is needed to evaluate (\ref{spin}), one can write down,
\begin{equation}\label{gr}
\delta G^<(t_0,t_0)=\frac{i}{2}\int\limits_{-\infty}^\infty\frac{d\varepsilon}{2\pi}\int\frac{d^2p}{(2\pi)^2}\Big\{g^R_{\bm{p}}\sigma_\mu\frac{\partial g_{\bm{p}}^<}{\partial\varepsilon}-\frac{\partial g_{\bm{p}}^R}{\partial\varepsilon}\sigma_\mu g^<_{\bm{p}}+g^<_{\bm{p}}\sigma_\mu\frac{\partial g_{\bm{p}}^A}{\partial\varepsilon}-\frac{\partial g_{\bm{p}}^<}{\partial\varepsilon}\sigma_\mu g^A_{\bm{p}}\Big\}\,\partial_t M_\mu,
\end{equation}
where $\mu=x,y,z$, while $g^R$, $g^A$, and $g^<$ are the bare retarded, advanced, and lesser Green's functions respectively. To derive the expression (\ref{gr}) we made use of Fourier transformation $g_{\bm{p}}=\int d(t_1-t_2)g_{\bm{p}}(t_1-t_2)\,e^{i\varepsilon(t_1-t_2)}$ and integration by parts.

To finally close up the derivation we employ the fluctuation-dissipation theorem according to which $g^<(\varepsilon)=[g^A(\varepsilon)-g^R(\varepsilon)]f(\varepsilon)$, where $f(\varepsilon)=[e^{\beta(\varepsilon-\mu)}+1]^{-1}$ stands for the Fermi-Dirac distribution with the Fermi energy $\mu$. Thus, nonequilibrium spin density of conduction electrons (\ref{spin}) within linear response theory is determined by $Q_{\mu\nu}=Q^{(1)}_{\mu\nu}+Q^{(2)}_{\mu\nu}$, where
\begin{equation}\label{q1} 
Q_{\mu\nu}^{(1)}=\frac{1}{4}\mathrm{Tr}\Big[\sigma_\mu\int\limits_{-\infty}^\infty\frac{d\varepsilon}{2\pi}\int\frac{d^2p}{(2\pi)^2}\Big\{\frac{\partial g_{\bm{p}}^R}{\partial\varepsilon}\sigma_\nu g^R_{\bm{p}}-g^R_{\bm{p}}\sigma_\nu\frac{\partial g_{\bm{p}}^R}{\partial\varepsilon}+g^A_{\bm{p}}\sigma_\nu\frac{\partial g_{\bm{p}}^A}{\partial\varepsilon}-\frac{\partial g_{\bm{p}}^A}{\partial\varepsilon}\sigma_\nu g^A_{\bm{p}}\Big\}f(\varepsilon)\Big],
\end{equation}
which involves the integration over the whole Fermi sea, and
\begin{equation}\label{q2}
Q_{\mu\nu}^{(2)}=\frac{1}{4}\mathrm{Tr}\Big[\sigma_\mu \int\limits_{-\infty}^\infty\frac{d\varepsilon}{2\pi}\left(-\frac{\partial f(\varepsilon)}{\partial \varepsilon}\right)\int \frac{d^2p}{(2\pi)^2}\Big\{g^R_{\bm{p}}\sigma_\nu g^R_{\bm{p}} +g^A_{\bm{p}}\sigma_\nu g^A_{\bm{p}}-2g^R_{\bm{p}}\sigma_\nu g^A_{\bm{p}}\Big\}\Big],
\end{equation}
which selects the integration in the vicinity of the Fermi level. Generally, the form of $Q_{\mu\nu}$ belongs to the class of Kubo-St\v{r}eda formula, and, in essence, represents the response to the external stimulus in the form of $\partial_tM_\nu$. We can immediately establish a quantitative agreement between the result given by Eq.~(\ref{q2}) and the previous studies within a Kubo formalism \cite{Brataas2008,Starikov2010,Bhattacharjee2012,Mankovsky2013,Ebert2015} which allow a direct estimation within the framework of disordered alloys. Formally, the expression (\ref{q1}) corresponds to the so-called St\v{r}eda contribution. Such a term was originally identified in Ref.~ \cite{Streda1982} when studying quantum-mechanical conductivity. Notably, in Eq.~(\ref{q1}) each term represents the product of either retarded or advanced Green’s functions. In this case the poles of the integrand function are positioned on the same side of imaginary plane, making disorder correction smaller in the weak disorder limit (see, e.g., Ref.~\cite{Sinitsyn2007}). Meanwhile, having no classical analog this contribution appears to be important enough when the spectrum of the system is gapped and the Fermi energy is placed exactly in the gap \cite{Streda1982}. It is worth mentioning that the contribution due to Eq.~(\ref{q1}) has never been discussed in this context before. In the meantime, Kubo-St\v{r}eda expression for the components of the Gilbert damping tensor has been addressed from the perspective of first-principles calculations \cite{Freimuth2015} and current-induced torques \cite{Freimuth2017}.

\section*{Results and discussion}

Let us apply the formalism developed in the previous section to a prototypical model: we work out the Gilbert damping tensor for a Rashba ferromagnet with the magnetization $\bm{m}=\bm{z}$ aligned along the $z$ axis. In the limit of weak disorder the Green's function of a bare system can be expressed as
\begin{equation}\label{grefun}
g^R_{\bm{p}}(\varepsilon)=\left(\varepsilon-H_0-\Sigma^R\right)^{-1}=\frac{\varepsilon-\varepsilon_{\bm{p}}+i\delta+\alpha(\bm{\sigma}\times\bm{p})_z+(\Delta+i\eta)\sigma_z}{(\varepsilon-\varepsilon_{\bm{p}}+i\delta)^2-\alpha^2p^2-(\Delta+i\eta)^2},
\end{equation}
where $\varepsilon_{\bm{p}}=p^2/(2m)$ is the electron kinetic energy. We put the self-energy $\Sigma^R$ due to scattering off scalar impurities into Eq.~(\ref{grefun}), which is determined from $\Sigma^R=-i(\delta-\eta\sigma_z)$ (see, e.g., Ref.~\cite{Ado2016}). In particular, for $|\varepsilon|>|\Delta|$ we can establish that $\delta=1/(2\tau)$ and $\eta=0$ in the weak disorder regime to the leading order.

Without loss of generality, in the following we restrict the discussion to the regime $\mu>|\Delta|$, which is typically satisfied with high accuracy in experiments. As previously discussed, the contribution owing to the Fermi sea, Eq.~(\ref{q1}), can in some cases be ignored, while doing the momentum integral in Eq.~(\ref{q2}) results in,
\begin{equation}\label{sigma}
\frac{1}{m\tau}\int\frac{d^2p}{(2\pi)^2}g^R_{\bm{p}}(\varepsilon)\bm{\sigma}g^A_{\bm{p}}(\varepsilon)=\frac{\Delta^2}{\Delta^2+2\varepsilon\rho}\bm{\sigma}+\frac{\Delta\delta}{\Delta^2+2\varepsilon\rho}\left(\bm{\sigma}\times\bm{z}\right)+\frac{\Delta^2-\varepsilon\rho}{\Delta^2+2\varepsilon\rho}(\bm{\sigma}\times\bm{z})\times\bm{z},
\end{equation}
where $\rho=m\alpha^2$. Thus, thanks to the factor of delta function $\delta(\varepsilon-\mu)=-\partial f(\varepsilon)/\partial\varepsilon$, to estimate $Q^{(2)}_{\mu\nu}$ at zero temperature one should put $\varepsilon=\mu$ in Eq.~(\ref{sigma}). As a result, we obtain,
\begin{equation}\label{bare}
Q^{(2)}_{\mu\nu}=-\frac{1}{4\pi}\frac{m}{\Delta^2+2\mu\rho }\left(\begin{array}{ccc}
2\tau\mu\rho & \Delta & 0 \\
-\Delta & 2\tau\mu\rho & 0 \\
0 & 0 & 2\tau\Delta^2
\end{array}\right).
\end{equation}
Meanwhile, to properly account the correlation functions which appear when averaging over disorder configuration one has to evaluate the so-called vertex corrections, which from a physical viewpoint makes a distinction between disorder averaged product of two Green's function, $\langle g^R\sigma_\nu g^A\rangle_\mathrm{dis}$, and the product of two disorder averaged Green's functions, $\langle g^R\rangle_\mathrm{dis}\sigma_\nu\langle g^A\rangle_\mathrm{dis}$, in Eq.~(\ref{q2}). Thus, we further proceed with identifying the vertex part by collecting the terms linear in $\delta$ exclusively,
\begin{equation}
\bm{\Gamma}^\sigma=A\bm{\sigma}+B(\bm{\sigma}\times\bm{z})+C(\bm{\sigma}\times\bm{z})\times\bm{z},
\end{equation}
provided $A=1+\Delta^2/(2\varepsilon\rho)$, $B=(\Delta^2+2\varepsilon\rho)\Delta\delta/(\Delta^2+\varepsilon\rho)^2$, and $C=\Delta^2/(2\varepsilon\rho)-\varepsilon\rho/(\Delta^2+\varepsilon\rho)$. To complete our derivation we should replace $\sigma_\nu$ in Eq.~(\ref{q2}) by $\Gamma^\sigma_\nu$ and with the aid of Eq.~(\ref{sigma}) we finally derive at $\varepsilon=\mu$,
\begin{equation}\label{vercor}
Q^{(2)}_{\mu\nu}=\left(\begin{array}{ccc}
Q_{xx} & Q_{xy} & 0 \\
-Q_{xy} & Q_{xx} \\
0 & 0 & -m\tau\Delta^2/(4\pi\mu\rho)
\end{array}\right).
\end{equation}
We defined $Q_{xx}=-m\tau\mu\rho/[2\pi(\Delta^2+\mu\rho)]$ and $Q_{xy}=-m\Delta(\Delta^2+2\mu\rho)/[4\pi(\Delta^2+\mu\rho)^2]$, which unambiguously reveals that account of vertex correction substantially modifies the results of the calculations. With the help of Eqs.~(\ref{spin}), (\ref{bare}), and (\ref{vercor}) we can write down LLG equation. Slight deviation from collinear configurations are determined by $x$ and $y$ components ($m_x$ and $m_y$ respectively, so that $|m_x|,|m_y|\ll1$). The expressions (\ref{bare}) and (\ref{vercor}) immediately suggest that the Gilbert damping at the interface is a scalar, $\alpha_G$,
\begin{equation}\label{fineq}
\partial_t\bm{m}=\tilde{\gamma}\bm{H}_\mathrm{eff}\times\bm{m}+\alpha_G\bm{m}\times\partial_t\bm{m},
\end{equation}
where the renormalized gyromagnetic ratio and the damping parameter are,
\begin{equation}
\tilde{\gamma}=\frac{\gamma}{1+\chi\Delta Q_{xy}}, \, \alpha_G=-\frac{\chi\Delta Q_{xx}}{1+\chi\Delta Q_{xy}}\approx-\chi\Delta Q_{xx}.
\end{equation}
In the latter case we make use of the fact that $m\chi\ll1$ for the NM thickness $d\sim$ 100 $\mu$m --- 100 nm. In Eq.~(\ref{fineq}) we have redefined the gyromagnetic ratio $\gamma$, but we might have renormalized the magnetization instead. From physical perspective, this implies the fraction of conduction electrons which become associated with the localized moment owing to $sd$ exchange interaction. With no vertex correction included one obtains
\begin{equation}\label{alp}
\alpha_G=\frac{m\chi}{2\pi\hbar}\frac{\tau\mu\rho\Delta}{\Delta^2+2\mu\rho},
\end{equation}
while taking account of vertex correction gives rise to a different result,
\begin{equation}\label{alpv}
\alpha_G=\frac{m\chi}{2\pi\hbar}\frac{\tau\mu\rho\Delta}{\Delta^2+\mu\rho}.
\end{equation}
To provide a quantitative estimate of how large the St\v{r}eda contribution in the weak disorder limit is, on condition that $\mu>|\Delta|$, we work out $Q_{\mu\nu}^{(1)}$. Using $\partial g^{R/A}(\varepsilon)/\partial\varepsilon=-[g^{R/A}(\varepsilon)]^2$ and the fact that trace is invariant under cyclic permuattaions we conclude that only off-diagonal components $\mu\neq\nu$ contribute. While the direct evaluation results in $Q_{xy}^{(1)}=3m\Delta/[2(\Delta^2+2\mu\rho)]$ in the clean limit. It has been demonstrated that including scattering rates $\delta$ and $\eta$ does not qualitatively change the results, leading to some smearing only \cite{Nunner2007}.

\begin{figure}
\includegraphics[scale=0.6]{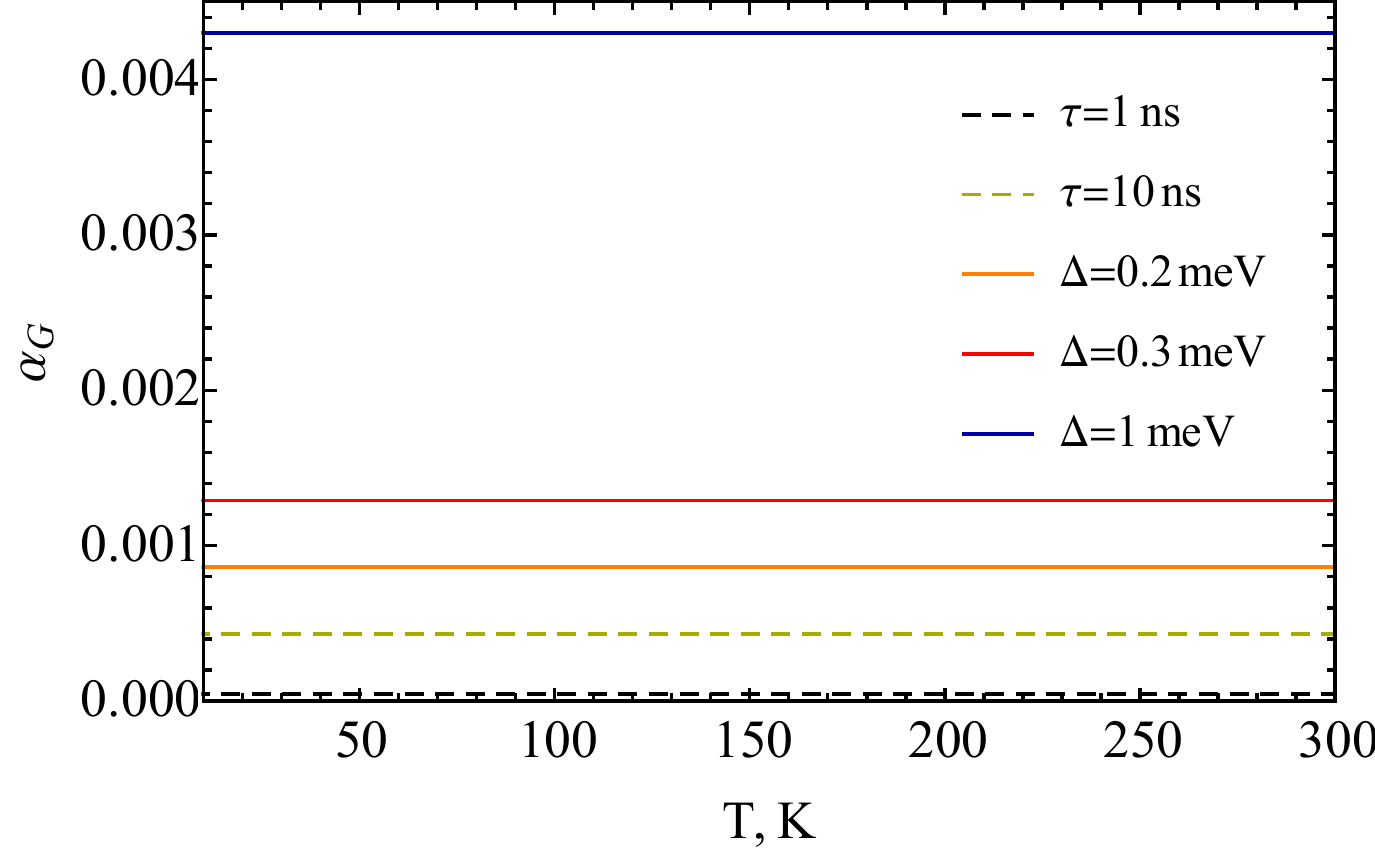}
\caption{Gilbert damping, obtained from numerical integration of Eq.~(\ref{q2}), shows almost no temperature dependence associated with thermal redistribution of conduction electrons. Dashed lines are plotted for $\Delta=1$ meV for $\tau=1$ and $\tau=10$ ns, whereas solid lines stand for $\Delta=0.2$, $0.3$, and $1$ meV for $\tau=100$ ns.}\label{fig:fig2}
\end{figure}

Interestingly, within the range of applicability of theory developed in this paper, the results of both Eqs.~(\ref{alp}) and (\ref{alpv}) depend linearly on scattering rate, being thus in qualitative agreement with the breathing Fermi surface model. Meanwhile, the latter does not yield any connection to the microscopic parameters (see, e.g., Ref. \cite{Eriksson2017} for more details). To provide with some quantitative estimations in our simulations we utilize the following set of parameters. Typically, experimental studies based on hyperfine field measurements equipped with DFT calculations \cite{Brooks1983} reveal the $sd$ Stoner interaction to be of the order of 0.2 eV, while the induced magnetization of $s$-derived states equals 0.002--0.05 (measured in the units of Bohr magneton, $\mu_B$). Thus, the parameter of $sd$ exchange splitting, appropriate for our model, is $\Delta\sim$ 0.2--1 meV. In addition, according to first-principles simulations we choose the Fermi energy $\mu\sim$ 3 eV. The results of numerical integration of (\ref{q2}) are presented in Fig.~\ref{fig:fig2} for several choices of $sd$ exchange and scattering rates, $\tau$. The calculations reveal almost no temperature dependence in the region up to room temperature for any choice of parameters, which is associated with the fact that the dominant contribution comes from the integration in a tiny region of the Fermi energy. Fig.~\ref{fig:fig2} also reveal a non-negligible dependence on the damping parameter with respect to both $\Delta$ and $\tau$, which illustrates that a tailored search for materials with specific damping parameter needs to address both the $sd$ exchange interaction as well as the scattering rate. From the theoretical perspective, the results shown in Fig.~\ref{fig:fig2} correspond to the case of non-interacting electrons with no electron-phonon coupling included. Thus, the thermal effects are accounted only via temperature-induced broadening which does not show up for $\mu>|\Delta|$.

\section*{Conclusions}

In this paper we proposed an alternative derivation of the Gilbert damping tensor within a generalized Kubo-St\v{r}eda formula. We established the contribution stemming from Eq.~(\ref{q1}) which was missing in the previous analysis within the linear response theory. In spite of being of the order of $(\mu\tau)^{-1}$ and, thus, negligible in the weak disorder limit developed in the paper, it should be properly worked out when dealing with more complicated systems, e.g., gapped materials such as iron garnets (certain half metallic Heusler compounds). For a model system, represented by a Rashba ferromagnet, we directly evaluated the Gilbert damping parameter and explored its behaviour associated with the temperature-dependent Fermi-Dirac distribution. In essence, the obtained results extend the previous studies within linear response theory and can be further utilized in first-principles calculations. We believe our results will be of interest in the rapidly growing fields of spintronics and magnonics.

\section*{Acknowledgements}

A.A.P. acknowledges the support from the Russian Science Foundation Project No. 18-72-00058. O.E. acknowledges support from eSSENCE, the Swedish Research Council (VR), the foundation for strategic research (SSF) and the Knut and Alice Wallenberg foundation (KAW). D.Y. acknowledges the support from the Russian Science Foundation Project No. 17-12-01359.

\section*{Author contributions statement}

D.Y. conceived the idea of the paper and contributed to the theory. A.A.P. wrote the main manuscript text, performed numerical analysis and prepared figures 1-2. M.I.B. and O.E. contributed to the theory. All authors reviewed the manuscript. 

\section*{Additional information}

{\bf Competing interests} The authors declare no competing  interests.

\end{document}